\documentclass[aps,IOP,onecolumn,superscriptaddress,groupedaddress]{revtex4-1}  
\usepackage{graphicx,amssymb}
\usepackage{color,ulem}
\usepackage{bm}   

\newcommand{\eqref}[1]{(\ref{#1})}

\usepackage{subfigure}
\usepackage{xcolor}
\usepackage{soul}
\soulregister\cite7
\soulregister\citep7
\soulregister\citet7
\soulregister\ref7
\soulregister\pageref7

\begin{document}
\title{Behavior of quantum coherence in the ultrastrong and deep strong coupling regimes of light-matter
 system}

 \author{Yu-qiang Liu$^{1}$}
 \email{Contact author: liuyuqiang@htu.edu.cn}
  \author{Qiulin Long$^{1}$}
\author{Yi-jia, Yang$^2$, Zheng Liu$^2$}
\author{Ting-ting, Ma$^3$}
\author{Bao-qing, Guo$^4$}
\author{Xingdong, Zhao$^1$}
\author{Zunlue, Zhu$^1$}
\author{Wuming, Liu$^5$}
\author{Chang-shui Yu$^{2}$}
\email{Contact author: ycs@dlut.edu.cn}
\affiliation{$^1$School of Physics, Henan Normal University, Xinxiang 453007, China}
\affiliation{$^2$School of Physics, Dalian University of Technology, Dalian 116024,
P.R. China}
\affiliation{$^3$Key Laboratory of Low-Dimensional Quantum Structures and Quantum Control of Ministry of Education, Department of Physics and Synergetic Innovation Center for Quantum Effects and Applications, Hunan Normal University, Changsha 410081, China}
\affiliation{$^4$Quantum Information Research Center, Shangrao Normal University, Shangrao 334001, China}
\affiliation{$^5$Beijing National Laboratory for Condensed Matter Physics, Institute of Physics, Chinese Academy of Sciences, Beijing 100190, People’s Republic of China}

\date{\today}

\begin{abstract}
The ultrastrong and deep strong coupling regimes exhibit a variety of intriguing physical phenomena. In this work, we utilize the Hopfield model of a two-mode bosonic system, with each mode interacts with a heat reservoir, to research the behavior of quantum coherence. Our results indicate that a coupled oscillator system can exhibit significant quantum coherence in the ultrastrong and deep strong coupling regimes. In the ground state, the photon-mode and the matter-mode coherences are equal. The larger coherences that encompass the photon mode, the matter mode, and the overall system are achieved at lower optical frequencies and with increased coupling strengths. Notably, the the beam-splitter and phase rotation terms alone does not generate coherences for either total coherence or subsystem coherences; instead, the generation of quantum coherences originates from the one-mode and two-mode squeezing terms. When heat environments are present, the total coherence can be enhanced by the the beam-splitter and phase rotation terms, while it has no effect on subsystem coherences. Moreover, when the one-mode and two-mode squeezing terms and the the beam-splitter and phase rotation terms are considered together, the total coherence increases with stronger coupling. We also observe that lower frequencies maximize total coherence in the deep strong coupling regime. These results demonstrate that the ultrastrong and deep strong coupling regimes give rise to novel characteristics of quantum coherence. This work provides valuable insights into the quantum coherence properties, particularly in the ultrastrong and deep strong coupling regimes between light and matter and may have potential applications in quantum information processing.
\end{abstract}

%\pacs{}

\maketitle
\section{Introduction}

Recently, significant research has been done on the ultrastrong and deep strong coupling regimes of light-matter interaction, where the coupling strengths are comparable to or even exceed the energies of the subsystems \cite{RevModPhys.91.025005, WOS:000542185800009, RevModPhys.93.025005, WOS:000599002700001, PhysRevApplied.20.024039, WOS:000416229300001, WOS:000558140400001, PhysRevA.84.043832}. When the coupling strength $g$ of light and matter is increased and reaches a considerable fraction of the transition frequency $\omega$ of the system, the ultrastrong $g/\omega>0.1$ or deep strong  $g/\omega>1$ coupling regimes can be realized \cite{RevModPhys.91.025005, WOS:000542185800009}.  Recent experimental advances in these coupling regimes have enhanced our understanding of light-matter interactions and the fundamental physics involved \cite{RevModPhys.91.025005, WOS:000542185800009, RevModPhys.93.041001, RevModPhys.93.025005, WOS:000599002700001, WOS:000416229300001, WOS:000558140400001, PhysRevApplied.20.024039}. 
Various quantum platforms have demonstrated the ultrastrong and deep strong coupling regimes, including cavity quantum electrodynamics (QED) \cite{RevModPhys.91.025005, WOS:000542185800009, PhysRevLett.124.040404}, circuit QED \cite{RevModPhys.93.025005, PhysRevLett.112.016401, PhysRevA.94.033827}, semiconductor quantum well systems \cite{PhysRevB.72.115303, WOS:000413057500067, PhysRevLett.105.196402}, and other hybrid quantum systems \cite{PhysRevB.97.024109, WOS:000554831500031, PhysRevApplied.16.034029}. A range of physical phenomena are being investigated in these regimes, including polariton detection \cite{PhysRevA.86.063831}, multiphoton quantum Rabi oscillations \cite{PhysRevA.92.063830}, quantum heat transport \cite{RevModPhys.93.041001, PhysRevE.107.044121, 10.1063/5.0160675}, Berry phase and topology \cite{PhysRevB.107.195104}, dynamical phase transitions \cite{PhysRevA.104.043712}, photon blockade \cite{PhysRevA.94.033827, PhysRevLett.109.193602}, pure dephasing \cite{PhysRevLett.130.123601}, and the inversion of qubit energy levels \cite{PhysRevLett.120.183601}. Currently, the ultrastrong and deep strong coupling regimes show considerable promise for a wide range of applications, including quantum simulation \cite{PhysRevLett.121.040505, PhysRevX.2.021007, PhysRevLett.105.263603}, quantum optical phenomena \cite{PhysRevLett.109.193602, PhysRevA.92.023842, PhysRevA.97.013851, WOS:000414915900027}, and quantum computation \cite{PhysRevLett.107.190402, PhysRevA.81.042311}. 
From these perspectives, the ultrastrong and deep strong coupling regimes in light–matter interactions have unveiled novel opportunities for exploring a wide array of physical phenomena.

Quantum resource theories provide a powerful mathematical framework for researching various diverse quantum resources and play an essential role in advancing quantum computation \cite{RevModPhys.91.025001}. In the era of quantum technology, quantum coherence has emerged as one of the most critical physical resources \cite{RevModPhys.89.041003}. Quantum coherence, which arises from the superposition of quantum states, is essential to a range of fields, including quantum thermodynamics \cite{PhysRevLett.132.200201, PhysRevE.102.062152, WOS:001163165600001, WOS:000352636000001}, quantum information processing \cite{PhysRevLett.116.160406}, quantum optics \cite{PhysRev.131.2766}, and quantum metrology \cite{doi:10.1126/science.1104149}. Research on quantum coherence encompasses both fundamental investigations into the formulation of resource theories and practical approaches for manipulating quantum coherence across a variety of systems. A comprehensive framework for quantifying quantum coherence was established in seminal work \cite{PhysRevLett.113.140401}. Subsequent studies have introduced additional measures based on the robustness of coherence \cite{PhysRevLett.116.150502}, trace-distance \cite{PhysRevA.93.012110}, relative entropy \cite{PhysRevA.93.032111}, quantum skew information \cite{PhysRevA.95.042337}, the geometric measure \cite{PhysRevLett.115.020403, PhysRevLett.119.140402}, pure-state coherence \cite{PhysRevA.101.062114}, Fisher information \cite{PhysRevA.106.052432}, among others. Many systems have been employed to investigate the characteristics and properties of quantum coherence. These include two-level systems \cite{WOS:001274844700001, PhysRevE.102.062152, PhysRevA.98.012102, PhysRevResearch.1.033097}, chains of qubits \cite{PhysRevA.107.012221}, cavity magnomechanical systems \cite{PhysRevB.109.064412}, coupled bosonic modes \cite{PhysRevA.106.032401}, and spinning magnomechanical systems \cite{PhysRevA.109.013719}. Recent studies have proposed that quantum coherence can be exploited for a variety of applications, including performing quantum thermodynamic tasks \cite{PhysRevResearch.1.033097, PhysRevLett.125.180603, PhysRevA.102.042220}, improving the efficiency of biomolecular switches \cite{PhysRevA.110.012411}, and enabling efficient quantum teleportation \cite{PhysRevA.108.042620}. Nonetheless, little attention has been devoted to investigating the behavior of quantum coherence within the ultrastrong and deep strong coupling regimes, as most current research primarily focuses on weak or normal light–matter interactions.

In this work, we go a step further to investigate the behavior of quantum coherence in the strong, ultrastrong and deep strong coupled light-matter system. To research the behavior of quantum coherence, we derive the covariance matrix for a two-mode full Hopfield model by employing the global master equation. We subsequently analyze the behavior of quantum coherence and the physical mechanisms based on the covariance matrix and coherence measure across the strong, ultrastrong, and deep strong coupling regimes. The structure of this work is as follows. In Sec. \ref{Sec. II}, we introduce the Hopfield model for two coupled bosonic modes and employ the master equation to analyze the system dynamics. In Sec. \ref{Sec. III}, we investigate ground-state quantum coherence and the impact of thermal effects on quantum coherence across the strong, ultrastrong, and deep strong coupling regimes. The conclusion is presented in Sec. \ref{conclusion}.

\section{Physical model and dynamics} \label{Sec. II}
%%%%%%%%%%%%%%%%%%%%%%%%%%%%%%%%%%%%%%%%%%%%%%%%%%%%%%%%%%%%%%%%%%%
\begin{figure}
\includegraphics[width=0.7\columnwidth]{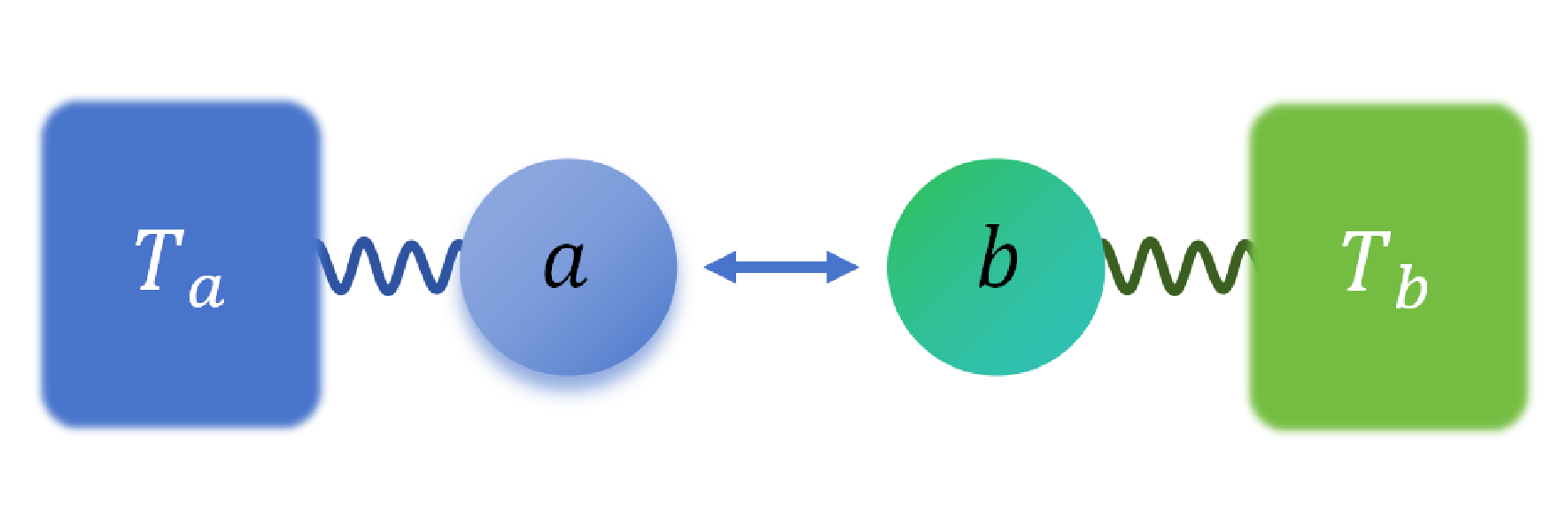}% Here is how to import EPS art
\caption{\label{fig:reservoir}
The illustration depicts a hybrid system of two coupled bosonic modes. The photon mode $a$ and the matter mode $b$ are in contact with a heat reservoir at different temperatures, $T_a$ and $T_b$, respectively. }
\end{figure}
%%%%%%%%%%%%%%%%%%%%%%%%%%%%%%%%%%%%%%%%%%%%%%%%%%%%%%
As depicted in Fig. \ref{fig:reservoir}, a two-mode bosonic systems version of the full Hopﬁeld Hamiltonian can read as \cite{WOS:000554831500031, PhysRevApplied.20.024039, PhysRevB.97.024109, WOS:000599002700001, PhysRevLett.105.196402} (In units of $\hbar=1$ and $k_{B}=1$)
\begin{equation} \label{H_{S}}
H_{\mathrm{Hop}} =\omega_{a} a^{\dagger} a+\omega_{b} b^{\dagger} b+\mathrm{i} g (a+a^{\dagger})(b-b^{\dagger})+D (a^{\dagger}+a)^2,
\end{equation}
where $g$ denotes the coupling strength of photon and matter, and $D$ is defined as $D = \frac{g^2}{\omega_{b}}$, referred to as the diamagnetic term. The interaction encapsulates two forms: squeezing and mix-mode coupling. The Hamiltonian Eq. (\ref{H_{S}}) can be rewritten as the following form \cite{PhysRevB.72.115303}
\begin{eqnarray}
&H_{\mathrm{Hop}}=H_{0}+H_{\mathrm{res}}+H_{\mathrm{anti}},\\
&H_{0}=\omega_a a^{\dagger} a+\omega_b b^{\dagger} b, \\
&H_{\mathrm{res}}=\mathrm{i} g(a^{\dagger} b- a b^{\dagger})+D(a^{\dagger} a+ a a^{\dagger}),\\
&H_{\mathrm{anti}}=\mathrm{i} g(a b- a^{\dagger} b^{\dagger})+D(a a+a^{\dagger} a^{\dagger}),
\end{eqnarray}
where $H_0$, $H_{\mathrm{res}}$, and $H_{\mathrm{_{anti}}}$ represent the terms of free Hamiltonian, the beam-splitter and phase rotation, and one-mode and two-mode squeezing, respectively \cite{RevModPhys.84.621}.
 A variety of quantum phenomena, including dissipation and the detection of polaritons \cite{PhysRevA.86.063831}, the breakdown of the Purcell effect \cite{PhysRevLett.112.016401}, and the quantum estimation of the diamagnetic term \cite{rossi2017probing}, have been analyzed in the framework of the Hopfield-type model. By employing the Hopfield transformation \cite{PhysRev.112.1555}, the Hamiltonian can be diagonalized as follows
\begin{equation}
H_{\mathrm{Hop}}^{\prime}=\sum_{j=\pm} \omega_{j} p_j^{\dagger} p_j,
\end{equation}
where $\omega_{j}$ denote the eigenfrequencies with 
$\omega_{\pm}=\sqrt{\frac{\omega_{b}^2+\omega^{2}_{C}+\omega^{2}_{a} \pm \sqrt{\left(\omega_{b}^2+\omega^{2}_{C}+\omega^{2}_{a}\right)^2-4 \omega^{2}_{a} \omega_{b}^2}}{2}}$ and $\omega_{C}=2g \sqrt{\frac{\omega_{a}}{\omega_{b}}}$. Moreover, the polariton operators are $p_{\pm}=w_{\pm} a+x_{\pm} b+y_{\pm} a^{\dagger}+z_{\pm} b^{\dagger}$ and satisfy the canonical commutation relation  $[p_{j}, p^{\dagger}_{i}]=\delta_{j, i}$. Additionally, the normalized coefficients are represented by the vector $\vec{p}_{\pm}=\lbrace w_{\pm}, x_{\pm}, y_{\pm}, z_{\pm}\rbrace$. 
These coefficients can be solved as \cite{PhysRev.112.1555}
\begin{eqnarray} 
\vec{p}_{\pm}=\pm \frac{1}{\sqrt{N}_{\pm}}\left(\begin{array}{c}
{\left[\frac{\omega_{\pm}^2}{\omega_{b}^2}-1\right] \frac{\omega_{\pm}+\omega_{a}}{2 \omega_{b}} \sqrt{\frac{\omega_{b}}{\omega_{a}}}}   \\
\mathrm{i} \sqrt{\frac{g^{2} \omega_{a}}{\omega^{3}_{b}}}\left(1+\frac{\omega_{\pm}}{\omega_{b}}\right)   \\
{\left[\frac{\omega_{\pm}^2}{\omega_{b}^2}-1\right] \frac{\omega_{\pm}-\omega_{a}}{2 \omega_{b}} \sqrt{\frac{\omega_{b}}{\omega_{a}}}}  \\
\mathrm{i} \sqrt{\frac{g^{2} \omega_{a}}{\omega^{3}_{b}}})\left(1-\frac{\omega_{\pm }}{\omega_{b}}\right)
\end{array}\right),
\end{eqnarray}
where the normalized coefficients $N_{\pm}=\frac{\omega_{\pm}}{\omega_{b}}[(1-\frac{\omega_{\pm}^2}{\omega_{b}^2})^2+ \frac{4 g^{2} \omega_{a}}{\omega^{3}_{b}} ]$. Similarly, the operators $a$, $b$ and their conjugate terms can also be represented by the polariton operators as
\begin{equation} \label{original operator}
\left(\begin{array}{c}
a \\
b \\
a^{\dagger} \\
b^{\dagger}
\end{array}\right)=\left(\begin{array}{cccc}
w_-^* & w_+^* & -y_- & -y_+ \\
x_-^* & x_+^* & -z_- & -z_+ \\
-y_-^* & -y_+^* & w_- & w_+ \\
-z_-^* & -z_+^* & x_- & x_+
\end{array}\right)\left(\begin{array}{c}
p_- \\
p_+ \\
p_-^{\dagger} \\
p_+^{\dagger}
\end{array}\right).
\end{equation}

\subsection{The interaction of system and thermal reservoir}

In general, the environment can be modelled by an ensemble of individual harmonic oscillators, as described in Ref. \cite{RevModPhys.59.1},
\begin{equation}
H_{R}=\sum_{\nu}H_{\nu}=\sum_{\nu, l}\omega_{\nu l}{c}_{\nu l}^{\dagger}{c}_{\nu l}, \, \nu=a, b,
\end{equation}
where $\omega_{\nu l}$, $c_{\nu l}$, and $c^{\dagger}_{\nu l}$ denote the frequencies, the annihilation and creation operators for the reservoir modes, respectively. Moreover, these operators satisfy the commutation relation $[c_{\nu l}, c^{\dagger}_{\nu^{\prime} l^{\prime}}]=\delta_{\nu \nu^{\prime}} \delta_{l l^{\prime}}$. The interaction of the system and heat reservoirs is given by
\begin{equation} \label{S-R}
H_{I}=(a+a^{\dagger}) (C^{\dagger}_{a}+C_{a})+(b+b^{\dagger}) (C^{\dagger}_{b}+C_{b}),
\end{equation}
with $C_{\nu}=\sum_{l}\kappa_{\nu l} c_{\nu l}$ and $\kappa_{\nu l}$ representing the system-reservoir coupling strength.
Employing Eq. (\ref{original operator}), the system-reservoir interaction can be expressed as
\begin{equation}  \label{S-R-r}
H_{I}^{\prime}=(C_{-}^{\dagger} p_{-}+p_{-}^{\dagger} C_{-})+ (B_{+}^{\dagger} p_{+}+p_{+}^{\dagger} B_{+}),
\end{equation}
where
$C_j=\mathcal{W}_{j}(C_a+C_a^{\dagger})+\mathcal{X}_{j}( C_b + C_b^{\dagger})
$ together with $\mathcal{W}_{j}=(w_j-y_j)$ and $\mathcal{X}_{j}=(x_j -z_j)$. 
In the ultrastrong coupling regime of light and matter, the counter-rotating terms of the system and reservoir interaction, as shown in Eqs. (\ref{S-R}) or (\ref{S-R-r}) should be neglected, as demonstrated in Refs. \cite{PhysRevA.84.043832, PhysRevA.89.023817}. Consequently, the Hamiltonian can be written in the form, 
\begin{equation} \label{S-R-RWA}
\tilde{H_{I}}=\sum_{j} [(\mathcal{W}_{j} C_{a}+\mathcal{X}_{j} C_{b}) p_{j}^{\dagger}+(\mathcal{W}^{*}_{j} C_{a}^{\dagger}+\mathcal{X}^{*}_{j} C_{b}^{\dagger}) p_{j}].
\end{equation}
Thus, the total Hamiltonian, which incorporates both system and environment, is given by 
\begin{equation}
H_{\mathrm{tot}}=H_{\mathrm{Hop}}^{\prime}+H_{R}+\tilde{H_{I}}.
\end{equation}

\subsection{Quantum master equation for open system} 
In the study of realistic dissipative systems, the master equation approach can be utilized by tracing out the degrees of freedom of the reservoirs to effectively describe the dynamics of the open system. The global master equation is formulated under the Born-Markov-secular approximation and is expressed in the Gorini-Kossakowski-Sudarshan-Lindblad form as follows \cite{breuer2002theory, PhysRevA.84.043832},
\begin{equation}
 \label{Eq:me}
\frac{d \rho}{dt} =\sum_{j=\pm}\lbrace
\Gamma(\omega_j) \mathcal{L}[p^{\dagger}_{j}] \rho+\Gamma(-\omega_j)\mathcal{L}[p_{j}] \rho \rbrace,
\end{equation}
where $\mathcal{L}[o] \rho$ denotes the dissipator, defined by $\mathcal{L}[o] \rho=o \rho o^{\dagger}-\frac{1}{2}\left\{o^{\dagger} o, \rho\right\}$ and $\Gamma(\pm \omega_j)=(J_{a} (\pm\omega_{j}) |\mathcal{W}_{j}|^{2}+J_{b} (\pm\omega_{j}) |\mathcal{X}_{j}|^{2} )$. The spectral densities are characterized by $J_{\nu} (\omega)=\zeta^{\nu}(\omega) N^{\nu}(\omega)$, $J_{\nu} (-\omega)=\mathrm{e}^{\frac{\omega}{T}} J_{\nu}(\omega)$, where the thermal occupation number is expressed as $N^{\nu} (\omega_{j})=\frac{1}{e^{ \omega_{j}/ T_{\nu}}-1}$. In this analysis, the Ohmic reservoir spectral response functions are assumed to be $\zeta^{a}(\omega) =\gamma \omega$ and $\zeta^{b}(\omega) =\kappa \omega$ for $\omega\geq 0$. 
For the oscillating terms $\mathrm{e}^{\mathrm{i} (\omega^{\prime}-\omega) t }$ at frequencies $|\omega-\omega^{\prime}|=\lbrace 0, 2 \omega_{+}, 2 \omega_{-}, \omega_{+} -\omega_{-}, 2 \omega_{-}, \omega_{+} +\omega_{-} \rbrace$, the secular approximation is justified provided that $(\omega_{+} -\omega_{-})\gg \lbrace\gamma, \kappa \rbrace$. Under this condition, the global master equation Eq. (\ref{Eq:me}) is well substantiated.  

\subsection{The measure of quantum coherence} 

A continuous-variable system comprising $N$ bosons is described by the mode operators $a_n$, with $n=1,..., N$, which satisfy the commutation relations $[a_{n}, a_{n^{\prime}}^{\dagger}]=\delta_{n, n^{\prime}}$. These $N$ bosons can be associated with the Hilbert space of the system $\mathcal{H}^{\otimes N}=\otimes_{n=1}^{N}\mathcal{H}_{n}$, corresponding to $N$ pairs of bosonic field operators $\lbrace a_{n}, a_{n}^{\dagger}\rbrace_{n=1}^{N}$. The free Hamiltonian of the system for the $n$th mode is given by $H_{n}=a_{n}^{\dagger} a_{n}$. It is convenient to define Hermitian quadrature operators $X_{n}=\frac{1}{\sqrt{2}}(a_{n}+a_{n}^{\dagger})$ and $P_{n}=\frac{\mathrm{i}}{\sqrt{2}} (a_{n}^{\dagger}-a_{n})$. The first moment vector is denoted as $\left\langle \vec{\xi} \right\rangle$, where $\vec{\xi}=(X_{1}, P_{1},..., X_{N}, P_{N})$. The quadrature variables $\vec{\xi}$ adhere to the commutation relation $[\xi_{n}, \xi_{m}]=\mathrm{i} \Omega_{n m}$, where $\Omega_{n m}$ is the matrix element of a $2N \times 2N$ symplectic matrix 
\begin{eqnarray} 
\mathbf{\Omega} = \bigoplus_{n=1}^{N} \left(
\begin{array}{cccc}
0& 1\\-1 & 0
\end{array}
\right).
\end{eqnarray}
The elements of the covariance matrix can be defined as $\sigma_{n, m}=\langle \lbrace \Delta \xi_{n}, \Delta \xi_{m}\rbrace\rangle=\langle \frac{\Delta \xi_{n} \Delta \xi_{m}+\Delta \xi_{m} \xi_{n}}{2}\rangle$
with $\Delta \xi_{n}=\xi_{n}-\langle \xi_{n}\rangle$ and $\langle \vec{\xi} \rangle=\mathrm{Tr} (\vec{\xi} \rho)$ \cite{RevModPhys.84.621}. A particular class of states is the Gaussian states, which can be completely characterized by the first and second moments \cite{1055441}.

For a one-mode Gaussian state $\rho \left(\sigma_{n}, \vec{\xi}_{n} \right)$ characterized by the covariance matrix 
\begin{eqnarray}
\left(
\begin{array}{cccc}
\sigma^{n}_{XX}& \sigma^{n}_{XP}\\
\sigma^{n}_{PX} & \sigma^{n}_{PP}
\end{array}
\right),
\end{eqnarray}
 and displacement vector $\vec{\xi}_{n}$, the quantum coherence of this state can be quantified by the relative entropy according to \cite{PhysRevLett.113.140401, PhysRevA.93.032111}
\begin{equation} \label{eq-c-one}
C_{n}[\rho \left(\sigma_{n}, \vec{\xi}_{n} \right)]=-f(v_{n})+f(\mu_{n}+\frac{1}{2}),
\end{equation}
where $f(x)=(x+\frac{1}{2}) \ln (x+\frac{1}{2})-(x-\frac{1}{2}) \ln (x-\frac{1}{2})$,$v_{n}=\sqrt{\mathrm{Det}(\sigma_{n})}$ and $\mu_{n}=\frac{\mathrm{Tr}(\sigma_{n})+\langle X_{n}\rangle^2 +\langle P_{n}\rangle^2-1}{2}$. In this context, the index $n$ may represent either $n=a$ or $n=b$.

Similarly, the quantum coherence for a two-mode Gaussian state is given by \cite{PhysRevLett.113.140401, PhysRevA.93.032111} 
\begin{equation} \label{eq-c-two}
C_{\mathrm{tot}}[\rho( \sigma, \vec{\xi} )]=S(\rho_{\mathrm{th}})-S(\rho),
\end{equation}
where $\sigma$, $\vec{\xi}$, $\rho_{\mathrm{th}}$ denote the covariance matrix, displacement vector, and reference thermal state, respectively. 
The von Neumann entropy $S(\rho)$ associated with the density operator $\rho$, is expressed as \cite{PhysRevA.59.1820}
\begin{equation}
S(\rho)=\sum_{l=1, 2} f(v_{l}),
\end{equation}
where $\lbrace v_{l}\rbrace_{l=1, 2}$, representing the symplectic eigenvalues of the two-mode system's covariance matrix $\sigma$. The entropy of the reference thermal state $S(\rho_{\mathrm{th}})$ is characterized by  
\begin{equation}
S(\rho_{\mathrm{th}})=\sum_{n=a, b} f(\mu_{n}+1/2).
\end{equation}
Higher values of $C_{n}$ and $C_{\mathrm{tot}}$ indicate greater quantum coherence in the subsystems and the total system, respectively.

\subsection{The covariance matrix of the hybrid system} 
%%%%%%%%%%%%%%%%%%%%%%%%%%%%%%%
\begin{figure}
\includegraphics[width=0.7\columnwidth]{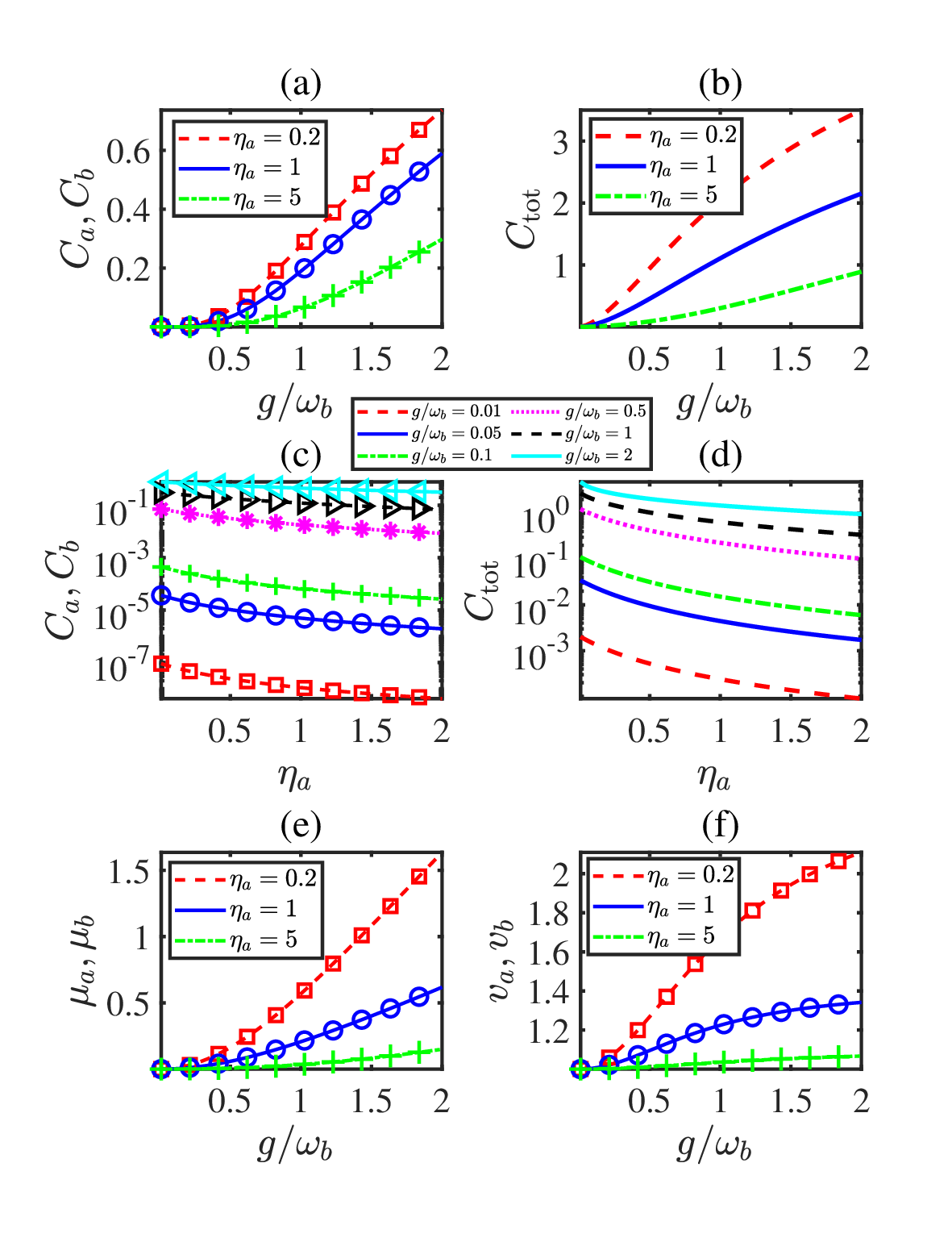}
\caption{\label{fig:C_gs}
Quantum coherence as a function of coupling strength (a-b) $g$ for different $\eta_a$, and frequency (c-d) $\eta_a$ for various coupling strength $g$. In (a), (b), (e) and (f), the lines (red dashed, blue solid, and green dash-dotted lines) give the coherences $C_a$, $C_{\mathrm{tot}}$ and ground-state average occupations $\mu_a$ and the symplectic eigenvalue $v_{a}$ of photon mode $a$ while the markers (red squares, blue circles, and green plus signs) correspond to the coherences $C_b$, mean occupations $\mu_b$ and symplectic eigenvalue $v_{b}$ of matter mode $b$. In (c), and (d), the lines (red dashed, blue solid, green dash-dotted, magenta dotted, black dashed, and cyan solid lines correspond to the coherences $C_a$ and $C_{\mathrm{tot}}$, while the markers (red squares, blue circles, green plus signs, magenta asterisks, black right-pointing triangles and cyan left-pointing triangles) give the coherences $C_b$. Note that $\eta_{a}=1$ denotes the resonant case and $\gamma=\kappa=10^{-3}$. For simplicity, all the parameters are expressed in units of frequency $\omega_b$ of the matter mode. }
\end{figure}
%%%%%%%%%%%%%%%%%%%%%%%%%%%%%%%
In phase space, the quadrature operators within the polariton basis are defined as $X_j=\frac{p_{j}+p_{j}^{\dagger}}{\sqrt{2}}$, $P_{j}=\mathrm{i} \frac{p_{j}^{\dagger}-p_{j}}{\sqrt{2}} $. As the evolution is nonunitary, hence to solve the covariance matrix, one needs obtain the dynamics of the second moments $\left\langle X_j^2\right\rangle (t)$, $\left\langle P_j^2\right\rangle (t)$, and $\left\langle X_j P_j\right\rangle (t)$. The remaining tasks involve determining the dynamics of $\langle p_{j}^2 \rangle (t)$, $\langle p_{j}^{\dagger 2} \rangle (t)$, $\langle p_{j}^{\dagger} p_{j} \rangle (t)$, $\langle p_{+}^{\dagger} p_{-} \rangle (t)$, $\langle p_{-}^{\dagger} p_{+}\rangle (t)$, $\langle p_{+}^{\dagger} p^{\dagger}_{-}\rangle (t)$, $\langle p_{+} p_{-}\rangle (t)$ and other six are dependent on the commutation relations. Following the master equation presented in Eq. (\ref{Eq:me}), the dynamics of $\langle p_{j}^2 \rangle$, $\langle p_{j}^{\dagger 2} \rangle$, $\langle p_{j}^{\dagger} p_{j} \rangle$, $\langle p_{+}^{\dagger} p_{-} \rangle$, $\langle p_{-}^{\dagger} p_{+}\rangle$, $\langle p_{+}^{\dagger} p^{\dagger}_{-}\rangle$, $\langle p_{+} p_{-}\rangle$ can be written as the following form, 
\begin{eqnarray} 
 \nonumber
\frac{\mathrm{d} \langle p_{j}^2 \rangle }{\mathrm{d} t}=&[-(\zeta^{a}(\omega_j) |\mathcal{W}_{j}|^2+\zeta^{b}(\omega_j)|\mathcal{X}_{j}|^2)-2 \mathrm{i} \omega_{j}] \langle p_{j}^2 \rangle, \\  \nonumber
\frac{\mathrm{d} \langle p_{j}^{\dagger 2} \rangle}{\mathrm{d} t}=&[-(\zeta^{a}(\omega_j) |\mathcal{W}_{j}|^2+\zeta^{b}(\omega_j)|\mathcal{X}_{j}|^2)+2 \mathrm{i} \omega_{j}] \langle p_{j}^{\dagger 2} \rangle,\\  \nonumber
\frac{\mathrm{d} \langle p_{j}^{\dagger} p_{j} \rangle}{\mathrm{d} t}=&\Gamma(\omega_j)-(\zeta^{a}(\omega_j) |\mathcal{W}_{j}|^2+\zeta^{b}(\omega_j)|\mathcal{X}_{j}|^2) \langle p_{j}^{\dagger} p_{j} \rangle, \\  
\frac{\mathrm{d} \langle p_{+}^{\dagger} p_{-} \rangle}{\mathrm{d} t}=&[-\sum_{j}(\zeta^{a}(\omega_j) |\mathcal{W}_{j}|^2+\zeta^{b}(\omega_j)|\mathcal{X}_{j}|^2)+\mathrm{i} (\omega_{+}-\omega_{-})] \langle p^{\dagger}_{+} p_{-} \rangle,\\  \nonumber
\frac{\mathrm{d} \langle p_{-}^{\dagger} p_{+} \rangle}{\mathrm{d} t}=&[-\sum_{j}(\zeta^{a}(\omega_j) |\mathcal{W}_{j}|^2+\zeta^{b}(\omega_j)|\mathcal{X}_{j}|^2) -\mathrm{i} (\omega_{+}-\omega_{-})] \langle p^{\dagger}_{-} p_{+} \rangle,\\  \nonumber
\frac{\mathrm{d} \langle p_{+}^{\dagger} p^{\dagger}_{-} \rangle}{\mathrm{d} t}=&[-\sum_{j}(\zeta^{a}(\omega_j) |\mathcal{W}_{j}|^2+\zeta^{b}(\omega_j)|\mathcal{X}_{j}|^2)+\mathrm{i}(\omega_{+}+\omega_{-})] \langle p_{+}^{\dagger} p^{\dagger}_{-} \rangle,\\  \nonumber
\frac{\mathrm{d} \langle p_{+} p_{-} \rangle}{\mathrm{d} t}=&[-\sum_{j}(\zeta^{a}(\omega_j) |\mathcal{W}_{j}|^2+\zeta^{b}(\omega_j)|\mathcal{X}_{j}|^2)-\mathrm{i}(\omega_{+}+\omega_{-})] \langle p_{+} p_{-} \rangle.
\end{eqnarray}
 The steady-state second moments are independent of initial conditions, and 
the nonzero second moments can be solved as $\left\langle X_j^2\right\rangle  =\left\langle P_j^2\right\rangle=\frac{\zeta^{a}(\omega_j) \vert \mathcal{W}_{j}\vert^2 (1+2 N^{a}(\omega_j))+\zeta^{b}(\omega_j) \vert \mathcal{X}_{j}\vert^2 (1+2 N^{b}(\omega_j))}{2 (\zeta^{a}(\omega_j) \vert \mathcal{W}_{j}\vert^2+\zeta^{b}(\omega_j) \vert \mathcal{X}_{j}\vert^2)}$, and $\left\langle X_j P_j\right\rangle=\frac{\mathrm{i}}{2}$. Note that the average occupy numbers are $N^{a}(\omega_j)=N^{b}(\omega_j)$ and we express it as $N(\omega_j)$ for simplification when the temperatures of heat reservoirs are identical, i.e. $T_{a}=T_{b}=T$. It means that $\left\langle X_j^2\right\rangle (\infty)$, and $\left\langle P_j^2\right\rangle (\infty)$ are independent of $\zeta^{\nu}_{j}$, and we can express $\left\langle X_j^2\right\rangle (\infty)$, $\left\langle P_j^2\right\rangle (\infty)$ as $\left\langle X_j^2\right\rangle (\infty)=\left\langle P_j^2\right\rangle (\infty)=\frac{1+2 N(\omega_j)}{2}$. 
Consequently, the steady-state covariance matrix $\sigma^{\prime}$ may be expressed in the polariton basis as
\begin{eqnarray} \label{Cov Matrix}
\sigma^{\prime}=\left(\begin{array}{cccc}
\left\langle X_{+}^2\right\rangle (\infty) & 0 & 0 & 0 \\
0 & \left\langle X_{+}^2\right\rangle (\infty) & 0 & 0 \\
0 & 0 & \left\langle X_{-}^2\right\rangle (\infty) & 0 \\
0 & 0 & 0 & \left\langle X_{-}^2\right\rangle (\infty)
\end{array}\right).
\end{eqnarray}
Based on the covariance matrix $\sigma^{\prime}$ in Eq. (\ref{Cov Matrix}), one can obtain the covariance matrix $\sigma$ in the bare representation by the unitary matrix $U$. The form of transformation $U$ can read as
\begin{equation} \label{eq:S}
U=\left(\begin{array}{cccc}
\sqrt{\frac{\omega_{a} \tau_{+}}{\omega_{+}}} & 0 & -\sqrt{\frac{\omega_{a} \tau_{-}}{\omega_{-}}}  & 0 \\
0& \sqrt{\frac{\omega_{+} \tau_{+}}{\omega_{a}}} & 0 &-\sqrt{\frac{\omega_{-} \tau_{-}}{\omega_{a}}} \\
0 & \epsilon_+ & 0  & -\epsilon_- \\
-\lambda_{+}& 0 & \lambda_- & 0\end{array}\right),
\end{equation}
where $\tau_{j}=\frac{(\omega_{j}^2-\omega_{b}^2)^2}{\omega_C^2 \omega_{b}^2+(\omega_{j}^2-\omega_{b}^2)^2}$, $\lambda_{j}=\omega_C \omega_{b}^{\frac{3}{2}} \sqrt{\frac{\chi_{j}}{\omega_{j}}}$, and $\epsilon_{j}=\omega_C \sqrt{\omega_{j} \omega_{b} \chi_{j}}$ with $\chi_{j}=\frac{1}{\omega_C^2 \omega_{b}^2+(\omega_{j}^2-\omega_{b}^2)^2}$.

%%%%%%%%%%%%%%%%%%%%%%%%%%%%%%%%%%%%%%%%%%%%%%%%%%%%%%%%%%%%%%%%%%%%%%%%%%%%
\begin{figure}
\subfigure{
\includegraphics[width=0.7\columnwidth]{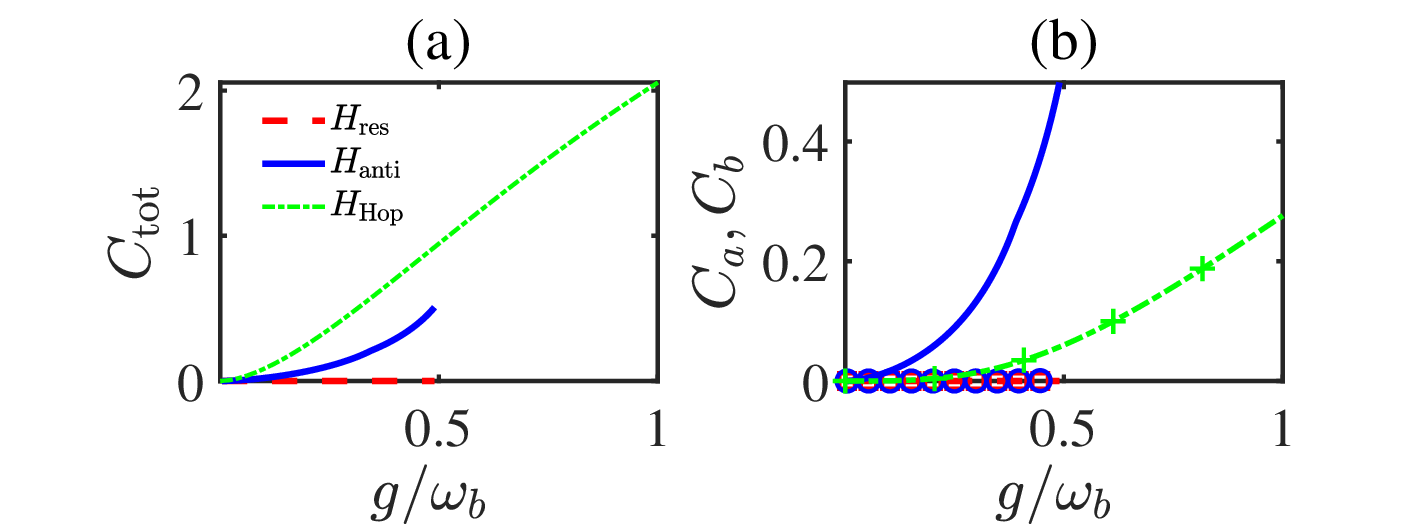}}
\caption{\label{fig:b}
Quantum coherences $C_{\mathrm{tot}}$ (a) and $C_{n}$ (b) versus the coupling strength $g$ in the ground state with different cases. The red dashed, blue solid, and green dash-dotted lines give the coherences $C_a$, $C_{\mathrm{tot}}$ (corresponding to the full Hamiltonian $H_{\mathrm{Hop}}$ without $H_{\mathrm{anti}}$ or without $H_{\mathrm{res}}$ or full Hamiltonian $H_{\mathrm{Hop}}$), while the markers (red squares, blue circles, and green plus signs) correspond to the coherences $C_b$. The other parameters are the same as in Fig. \ref{fig:C_gs}. }
\end{figure}
%%%%%%%%%%%%%%%%%%%%%%%%%%%%%%%%%%%%%%%%%%%%%%%%%%%%%%%%%%%%%%%%%%%%%%%%%%%%

In the bare representation, the covariance matrix in the ground state can read as
\begin{equation} \label{cm_gs}
\sigma_{gs}=\left(\begin{array}{cccc}
\frac{\omega_{a} \sum_{j}\frac{\tau_{j}}{\omega_{j}}}{2 } & 0 & 0 & \delta \\
0& \frac{\sum_j\omega_{j}\tau_{j}}{2\omega_{a}} & \delta& 0 \\
0 & \delta & \frac{\omega_{a} \sum_{j}\frac{\tau_{j}}{\omega_{j}}}{2}  & 0 \\
\delta & 0 & 0 & \frac{\sum_j\omega_{j}\tau_{j}}{2\omega_{a}} 
\end{array}\right),
\end{equation}
with $\delta=g \sum_{j} \omega_{j} \vartheta_{j}$, and $\vartheta_{j}=\frac{\tau_{j}}{(\omega_{j}^2-\omega_{b}^2)}$.
We can find that the covariance matrices $\sigma_a$ and $\sigma_{b}$ of the subsystems are identical.
The covariance matrix in the thermal environments can be expressed as 
\begin{eqnarray} \label{hamilton1}
\sigma = \left(
\begin{array}{cccc}
\sigma_a & \sigma_{ab}  \\
\sigma_{ab} ^{\mathrm{T}} & \sigma_b 
\end{array}
\right),
\end{eqnarray}
where the matrices $\sigma_a$, $\sigma_b$ and $\sigma_{ab}$ are, respectively,
\begin{eqnarray} \label{Cov-Matrix-a}
\sigma_{a}=\left(\begin{array}{cc}
\omega_{a} \sum_{j}\frac{\left\langle X_{j}^2\right\rangle (\infty)\tau_{j}}{\omega_j} & 0  \\
0 & \frac{\sum_j\omega_{j}\left\langle X_{j}^2\right\rangle (\infty)\tau_{j}}{\omega_{a}}  
\end{array}\right),
\end{eqnarray}
\begin{eqnarray} \label{Cov-Matrix-b}
\sigma_b= 4 g^2 \omega_{a}\left(\begin{array}{cc}
\sum_{j} \omega_{j}\left\langle X_{j}^2\right\rangle (\infty)\chi_{j} & 0  \\
0 & \omega_{b}^2\sum_{j} \left\langle X_{j}^2\right\rangle (\infty) \chi_{j} 
\end{array}\right),
\end{eqnarray}
and
\begin{eqnarray} \label{Cov-Matrix-c}
\sigma_{ab}=2\left(\begin{array}{cc}
0 & -g \omega_{a} \omega_{b} \sum_{j}  \frac{\left\langle X_{j}^2\right\rangle (\infty) \vartheta_{j}}{\omega_j} \\
g \sum_{j} \omega_{j} \left\langle X_{j}^2\right\rangle (\infty)\vartheta_{j} & 0
\end{array}\right).
\end{eqnarray}

\section{Results and discussion} \label{Sec. III}

%%%%%%%%%%%%%%%%%%%%%%%%%%%%%%%%%%%
\begin{figure}
\includegraphics[width=0.7\columnwidth]{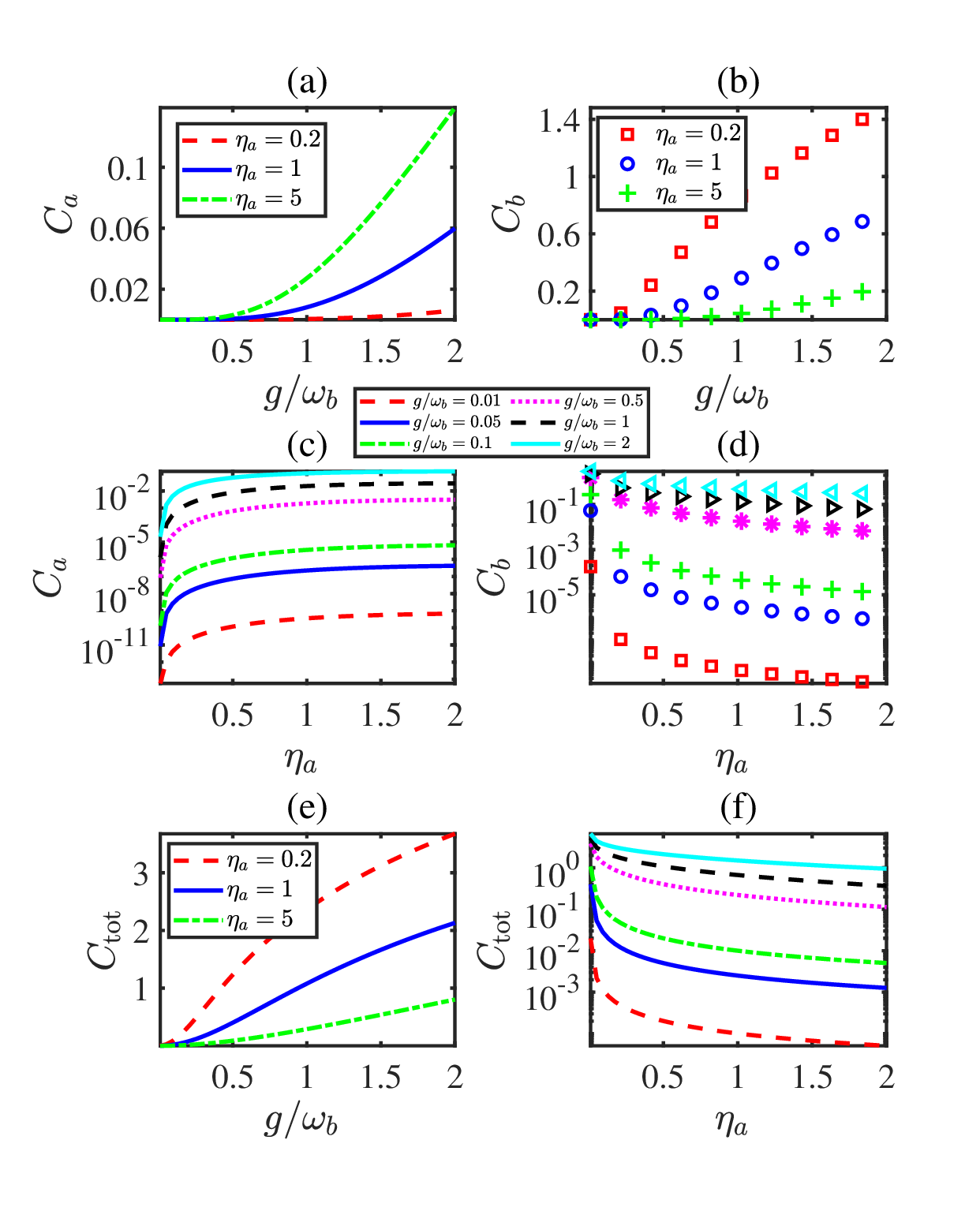}
\caption{\label{fig:Coh_local_g}
Quantum coherence as a function of coupling strength (a, b, e) $g$ for different $\eta_a$, and frequency (c, d, f) $\eta_a$ for various $g$. Note that $\eta_{a}=1$ denotes the resonant case. In (a), (b) and (e), the lines (red dashed, blue solid, and green dash-dotted lines) give the coherences $C_a$, $C_{\mathrm{tot}}$ and ground-state average occupations $\mu_a$ of photon mode $a$ while the markers (red squares, blue circles, and green plus signs) correspond to the coherences $C_b$ and ground-state mean occupations $\mu_b$ of matter mode $b$. In (c), (d) and (f), the lines (red dashed, blue solid, green dash-dotted, magenta dotted, black dashed, and cyan solid lines correspond to the coherences $C_a$ and $C_{\mathrm{tot}}$, while the markers (red squares, blue circles, green plus signs, magenta asterisks, black right-pointing triangles and cyan left-pointing triangles) give the coherence $C_b$. The parameters can take $T_{b}=T_{a}=T$ with $T=\omega_b$ and $\gamma=\kappa=10^{-3}$.}
\end{figure}
%%%%%%%%%%%%%%%%%%%%%%%%%%%%%%%%%%%%%

%%%%%%%%%%%%%%%%%%%%%%%%%%%%%%%%%%%%%%%%%%%%%%%%%%%%%%%%%%%%%%%%%%%%%%%%%%%%
\begin{figure}
\includegraphics[width=0.5\columnwidth]{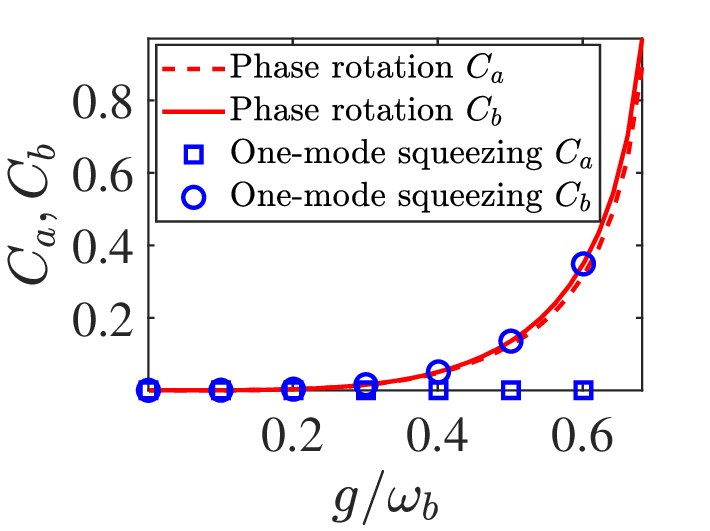}
\caption{\label{fig:C_No_A2}
Quantum coherence versus the coupling strength $g$ corresponding to the cases of the Hamiltonian only with one-mode squeezing $H_S+H_s$ or phase rotation $H_S+H_p$. The parameters can take $\eta_a=1$, $T_{a}=T_{b}=T$ with $T=\omega_b$, and $\gamma=\kappa=10^{-3}$. }
\end{figure}
%%%%%%%%%%%%%%%%%%%%%%%%%%%%%%%%%%%%%%%%%%%%%%%%%%%%%%%%%%%%%%%%%%%%%%%
We seek to explore quantum coherence within the strong, ultrastrong, and deep strong coupling regimes of two bosonic modes. Our primary focus is on both resonant and off-resonant cases within these coupling regimes. For clarity, we express the normalized polariton modes in terms of the matter frequency $\omega_b$ using the equation $\frac{\omega_{\pm}}{\omega_b}=\sqrt{\frac{1+\eta_{C}^2+\eta_{a}^2\pm\sqrt{(1 +\eta_{C}^2+\eta_{a}^2)^2-4 \eta_{a}^2}}{2}}$ with $\eta_{C}=\frac{\omega_C}{\omega_b}$ and $\eta_{a}=\frac{\omega_a}{\omega_b}$. For the sake of simplicity, all parameters in this discussion will be expressed in units of the matter mode frequency $\omega_b$. In Figs. \ref{fig:C_gs} (a) and \ref{fig:C_gs} (b), we observe that the cavity mode and matter mode demonstrate the same coherence, indicated as $C_a = C_b$. This conclusion can be drawn from the covariance matrix in Eq. (\ref{cm_gs}), where $\sigma_a$ and $\sigma_b$ are identical. For the smallest coupling, where $g/\omega_b \rightarrow 0$, there are no quantum coherences in both the subsystems and the total system. Moreover, for a fixed value of $\eta_{a}$, the coherences of the photon $C_a$, matter $C_b$, and total system $C_{\mathrm{tot}}$ gradually increase with stronger coupling strength. From Figs. \ref{fig:C_gs} (c) and \ref{fig:C_gs} (d), we observe that the coherences $C_a$, $C_b$ and $C_{\mathrm{tot}}$ decrease monotonically as the normalized optical frequency $\eta_a$ increases. The maximum coherence is attained at lower optical frequencies $\omega_a$. Moreover, both $C_a$ and $C_b$ increase as the coupling strengths rise from $g=0.01 \omega_{b}$, $0.05 \omega_{b}$, $0.1 \omega_{b}$, $0.5 \omega_{b}$, $\omega_{b}$ to $2 \omega_{b}$. As derived from Eqs. (\ref{eq-c-one}), (\ref{eq-c-two}), and (\ref{cm_gs}), and using the relation $p_j | G \rangle=0 $ in the polariton basis, the ground state $| G \rangle$ should be a squeezed state as shown in Ref. \cite{PhysRevB.72.115303}. Furthermore, according to the relation (\ref{original operator}), the occupation numbers can be solved as $\mu_{a}=\left\langle a^{\dagger} a \right\rangle=\sum_{j} |y_j|^2$, $\mu_{b}=\left\langle b^{\dagger} b \right\rangle=\sum_{j} |z_j|^2$. After some algebras, one can obtain $\mu_{a}=\mu_{b}=g^2 \omega_a(\frac{\tau_{-}}{\omega_{-}}+\frac{\tau_{+}}{\omega_{+}})$. In the ground state, $\mathrm{det}(\sigma)=\frac{1}{16}$ and $\mathrm{det} \sigma_a+\mathrm{det} \sigma_a+\mathrm{det} \sigma_{ab}=\frac{1}{2}$, the two symplectic eigenvalues are $1/2$. Hence the total coherence determines the mean occupations. This enhancement in coherence is attributed to the rising mean occupations that accompany stronger coupling strength, as illustrated in Fig. \ref{fig:C_gs} (e). Consequently, the ultrastrong and deep strong coupling regimes offer a viable avenue for achieving enhanced quantum coherences. Moreover, these figures illustrate that the total coherence $C_{\mathrm{tot}}$ exceeds the individual coherences of the photon and matter $C_a$ and $C_b$. This is because the reduced operation can decrease coherence. To analyze it, we plot the $v_a$ and $v_b$ as a function of coupling strength $g$ as shown in Fig. \ref{fig:C_gs} (d), and $v_n$ gradually increases with increasing coupling strength. According to Eq. (\ref{eq-c-two}), the coherences $C_n$ depend on the $f({\mu}_n)$ and $f(v_n)$. As $C_{\mathrm{tot}}=\sum_{n} f(\mu_n+\frac{1}{2})$, $C_{n}=-f(v_{n})+f(\mu_{n}+\frac{1}{2})$ with $f(v_n)>0$ and $f(\mu_n+\frac{1}{2})>0$, hence $C_{\mathrm{tot}}>C_n$. To give a intuitive understanding of generation of quantum coherences, we consider three cases: the full Hamiltonian $H_{\mathrm{Hop}}$ without $H_{\mathrm{anti}}$, the full Hamiltonian $H_{\mathrm{Hop}}$ without $H_{\mathrm{res}}$, and the full Hamiltonian $H_{\mathrm{Hop}}$. It can be seen from Fig. \ref{fig:b}, only the term $H_{\mathrm{res}}$ can not produce the coherences for $C_{\mathrm{tot}}$ and $C_{n}$. The generation of quantum coherences originate from the term $H_{\mathrm{anti}}$.

In the presence of heat reservoirs, we set the temperatures as $T_a = T_b = T$ with $T = \omega_b$. The analysis presented in Figs. \ref{fig:Coh_local_g} (a), \ref{fig:Coh_local_g} (b), and \ref{fig:Coh_local_g} (e) indicates that the quantum coherences $C_a$, $ C_b$, and the total coherence $C_{\mathrm{tot}}$, in particular $C_b$ and $C_{\mathrm{tot}}$ demonstrate an upward trend with increasing internal coupling strength, while maintaining a fixed value of $\eta_a$. The smaller values of $\eta_a$ lead to larger coherences $C_b$ and $ C_{\mathrm{tot}}$. However, the coherence $C_a$ weakens as $\eta_a$ varies from $\eta_a = 5, 1$ to $\eta_a = 0.2$. In Figs. \ref{fig:Coh_local_g} (c) and \ref{fig:Coh_local_g} (d), maximal coherence $C_a$ is achieved at larger $\eta_a$, while maximal $C_b$ occurs at smaller $\eta_a$. Moreover, from Figs. \ref{fig:Coh_local_g} (a), \ref{fig:Coh_local_g} (b), \ref{fig:Coh_local_g} (c), and \ref{fig:Coh_local_g} (d), it is also apparent that the subsystem coherences $C_a$ and $C_b$ differ significantly, with the matter coherence $C_b$ being larger than the photon coherence $C_a$.

\begin{figure}
\subfigure{
\includegraphics[width=0.8\columnwidth]{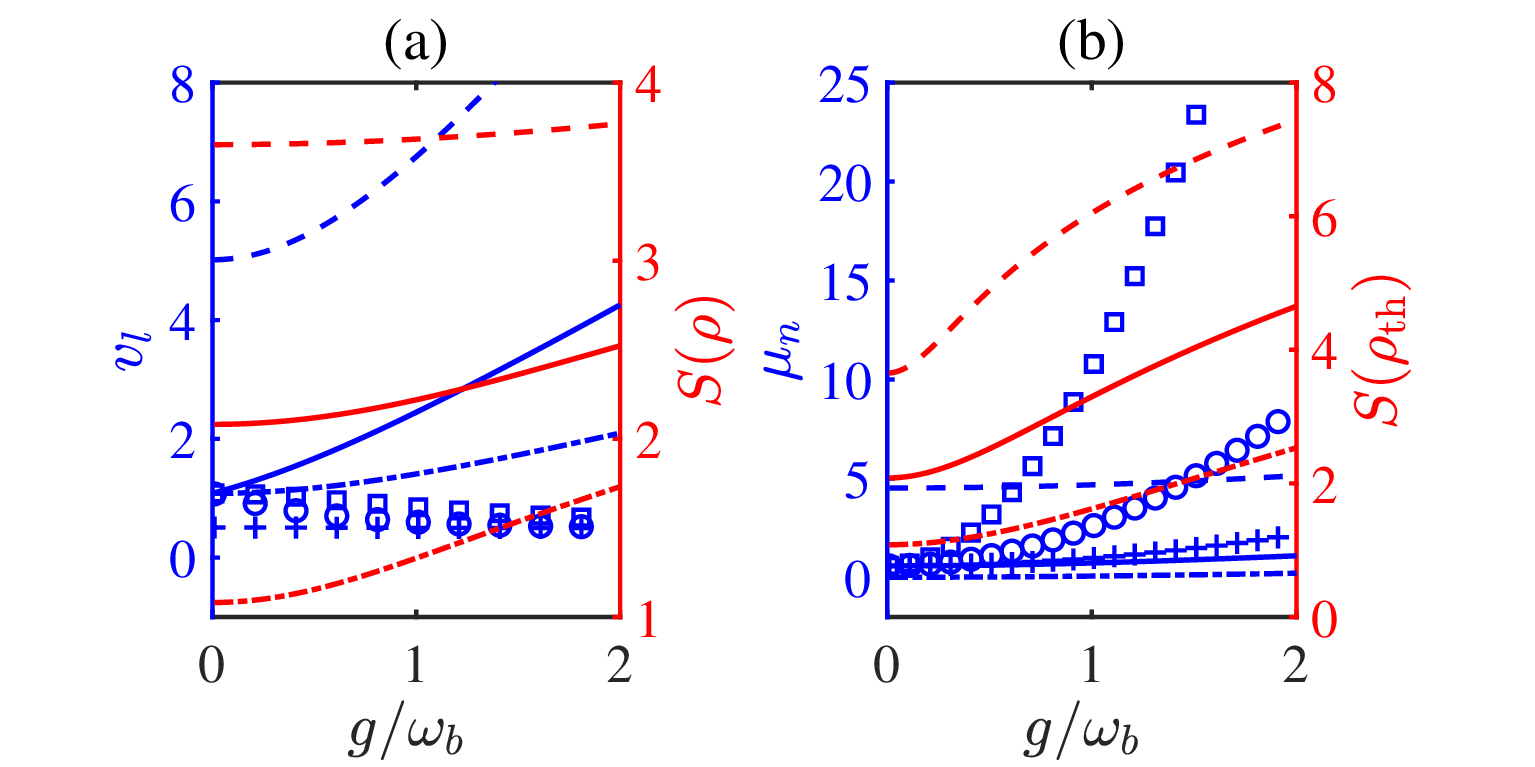}}
\caption{\label{fig:C_ana_local}
(a) The symplectic eigenvalues $v_{l}$ of covariance matrix $\sigma$ of two-mode system, the von Neumann entropy $S(\rho)$, and (b) the average occupations $\mu_n$, von Neumann entropy $S(\rho_{\mathrm{th}})$ versus the coupling strengths. The lines denote the symplectic eigenvalues $v_{1}$ and average occupation $\mu_a$ with $\eta_a=0.2, 1, 5$ (blue dashed, blue solid, and blue dash-dotted lines), and $v_{2}$ and $\mu_b$ (blue square, blue circle, and blue asterisk lines). The values of the parameter take $T=\omega_b$, and $\gamma=\kappa=10^{-3}$. }
\end{figure}

 This difference arises from the distinct forms of the covariance matrices for subsystems $a$ and $b$, as shown in Eqs. (\ref{Cov-Matrix-a}) and (\ref{Cov-Matrix-b}) in the presence of heat reservoirs. The total coherences are displayed in Figs. \ref{fig:Coh_local_g} (e) and \ref{fig:Coh_local_g} (f), indicating that quantum coherences can be enhanced in the ultrastrong and deep strong coupling regimes along with lower optical frequencies. The average occupation numbers of photon and matter modes can be solved as $\mu_{a}=\left\langle a^{\dagger} a \right\rangle=\sum_{j} [(|w_{j}|^2+ |y_{j}|^2) \left\langle p_{j}^{\dagger} p_{j}\right\rangle+|y_j|^2 ]$, $\mu_{b}=\left\langle b^{\dagger} b \right\rangle=\sum_{j} [(|x_{j}|^2+ |z_{j}|^2) \left\langle p_{j}^{\dagger} p_{j}\right\rangle +|z_j|^2]$, respectively. When the temperatures of heat reservoirs satisfy the condition $T_a=T_b=T$, one can express $\left\langle p_{j}^{\dagger} p_{j}\right\rangle$ as $\left\langle p_{j}^{\dagger} p_{j}\right\rangle=N(\omega_j)$. There exists a special case $D=0$ corresponding to the standard Dicke Hamiltonian, where the diamagnetic term is neglected \cite{WOS:000283646500011, WOS:000554831500031, PhysRevE.67.066203, PhysRevE.93.052118}. Hence, we explore the quantum coherence of two coupled bosonic modes without $D$ (further details are provided in Appendix \ref{Appendix A}). Based on this, one can also express the Hopfield model as $H_{\mathrm{Hop}}=H_{S}+H_{s}+H_{p}$ with $H_{S} =\omega_{a} a^{\dagger} a+\omega_{b} b^{\dagger} b+\mathrm{i} g (a+a^{\dagger})(b-b^{\dagger})$, the one-mode squeezing $H_{s}=D (a^{\dagger 2}+a^2)$ and the phase rotation $H_{p}=D(a^{\dagger} a+ a a^{\dagger})$ \cite{RevModPhys.84.621}. We also see that $C_b > C_a$ mainly originates from the contribution of the one-mode squeezing term. To elucidate the mechanism of coherence behavior, we present the symplectic eigenvalues $v_l$ of the covariance matrix $\sigma$, the mean occupations $\mu_{n}$, and their corresponding von Neumann entropy, as displayed in Fig. \ref{fig:C_ana_local}. From Fig. \ref{fig:C_ana_local} (a), it is evident that the symplectic eigenvalue $v_1$ increase gradually with coupling strength, while $v_2$ decreases. It is also shown that the symplectic eigenvalues $v_1$ are observed to be greater than $v_2$, and smaller values of $\eta_a$ are associated with higher $v_1$ and lower $v_2$. The function $f(v_1)$ grows faster with coupling strength than $f(v_2)$, which leads to the increasing von Neumann entropy $S(\rho)$. Fig. \ref{fig:C_ana_local} (b) reveals that the mean occupations for both $\mu_a$ and $\mu_b$ increase with stronger coupling strengths, leading to a significant enhancement of the von Neumann entropy $S(\rho_{\mathrm{th}})$. A comparative analysis of the values of $S(\rho)$ and $S(\rho_{\mathrm{th}})$ illustrates that both the ultrastrong and deep strong coupling regimes, and lower optical frequencies can show greater quantum coherences. 

%%%%%%%%%%%%%%%%%%%%%%%%%%%%%%%%%%%%%%%%%%%%%%%%
%%%%%%%%%%%%%%%%%%%%%%%%%%%%%%%%%%%%%%%%%%%%%%%%%
 %%%%%%%%%%%%%%%%%%%%%%%%%%%%%%%%%%%%%%%%%%%%%%%%%%%%
\begin{figure}
\includegraphics[width=0.7\columnwidth]{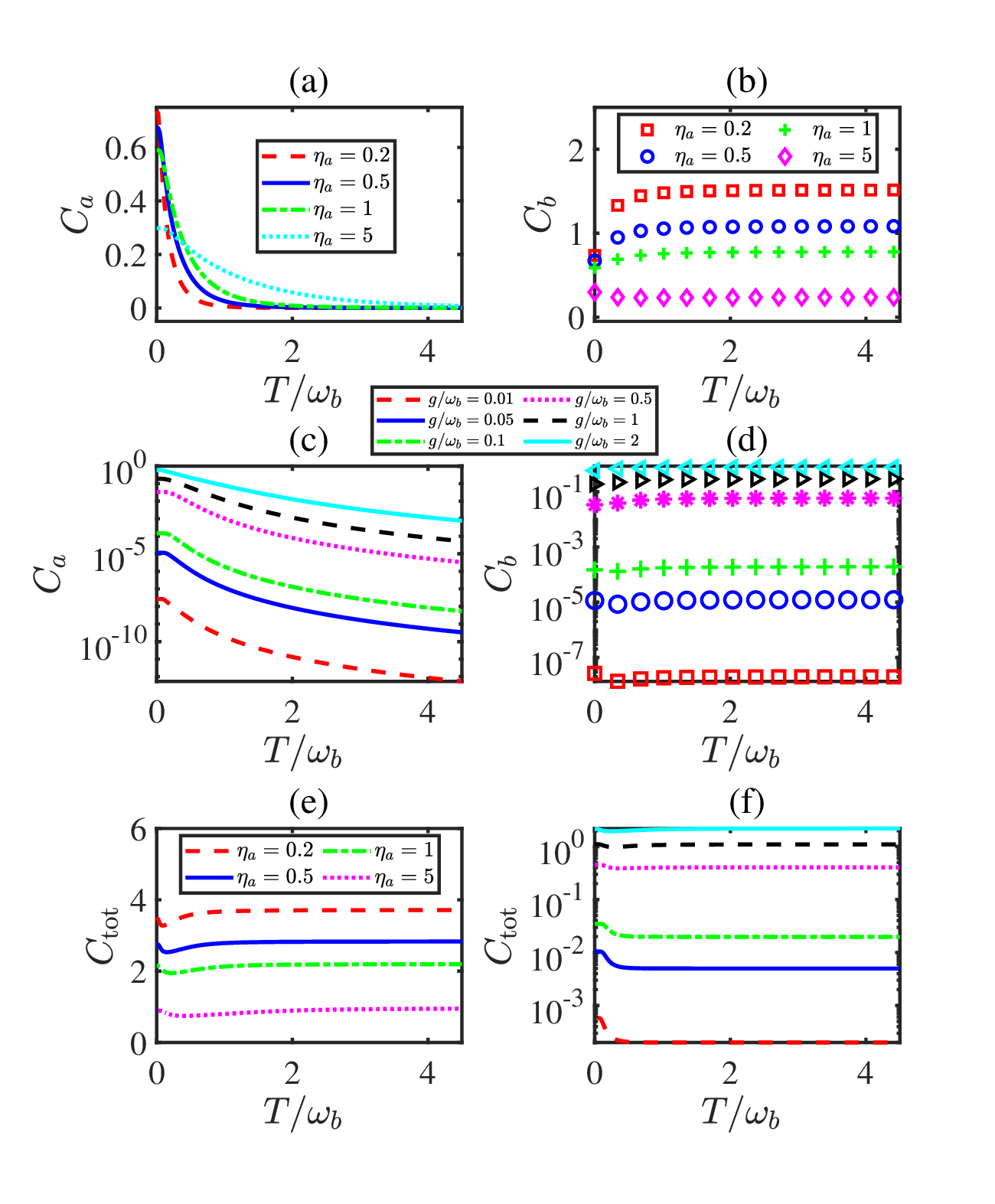}
\caption{\label{fig:C_local}
Quantum coherences as a function of temperature $T$ for (a, b, e) different optical frequencies and (c, d, f) different coupling strengths. In (a), (b), and (e), the lines give the coherences $C_a$, $C_{\mathrm{tot}}$, while the markers correspond to the coherences $C_b$ and ground-state mean occupations $\mu_b$ of matter mode $b$. In (c), (d) and (f), the lines correspond to the coherences $C_a$ and $C_{\mathrm{tot}}$, while the markers give the coherence $C_b$. The values of the parameter can take (a, b, e) with $g=2\omega_{b}$; (c, d, f) with $\eta_a=1$. Note that $\eta_{a}=1$ denotes the resonant case. In addition, the dissipation strengths are $\gamma=\kappa=10^{-3}$. }
\end{figure}
%%%%%%%%%%%%%%%%%%%%%%%%%%%%%%%%%%%%%%%%%%%%%%%%%%

%%%%%%%%%%%%%%%%%%%%%%%%%%%%%%%%%%%%%%%%%%%%%%%%%%%%%%%%%%%%%%%%%%%%%%%%%%%%%
\begin{figure}
\includegraphics[width=0.7\columnwidth]{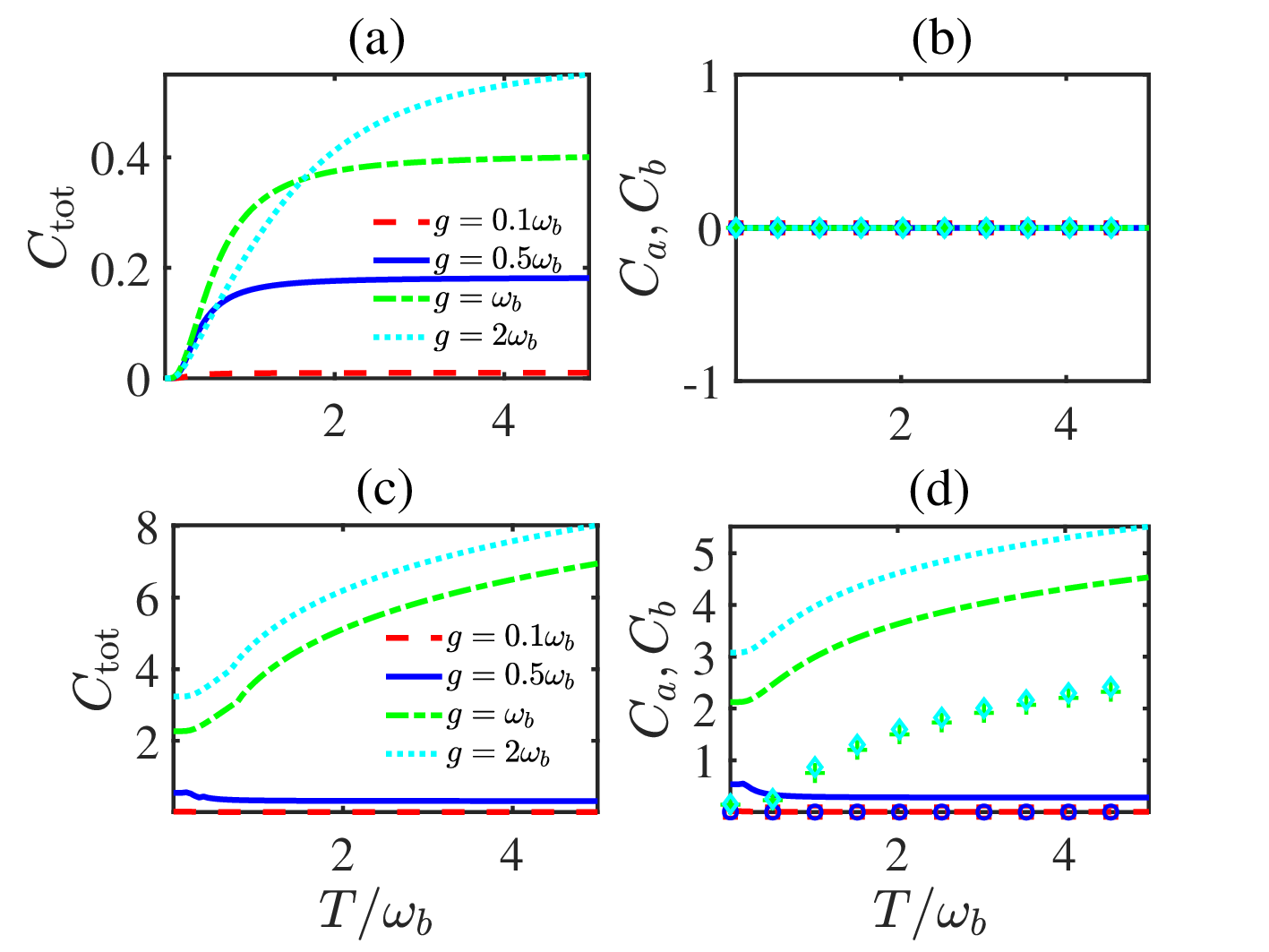}
\caption{\label{fig:bb}
Quantum coherences $C_{\mathrm{tot}}$ (a, c) and $C_n$ (b, d) versus the temperatures with different coupling strengths $g$. Note that $\eta_{a}=1$ denotes the resonant case. (a-b) and (c-d) correspond to the full Hamilton $H_{\mathrm{Hop}}$ without the $H_{\mathrm{anti}}$ and without the $H_{\mathrm{res}}$, respectively. Note that the lines and markers in (b) and (d) are subsystem coherences $C_a$ and $C_b$, respectively. The other parameters are the same as in Fig. \ref{fig:C_local}.  }
\end{figure}
%%%%%%%%%%%%%%%%%%%%%%%%%%%%%%%%%%%%%%%%%%%%%%%%%%%%%%%%%%%%%%%%%%%%%%%%%%%%%

Let us investigate the effect of thermal noise on quantum coherence under the assumption of no thermal bias $T_{a}=T_b=T$. As illustrated in Figs. \ref{fig:C_local} (a), \ref{fig:C_local} (b), \ref{fig:C_local} (c), and \ref{fig:C_local} (d), the coherences of the individual subsystems $C_a$ and $C_{b}$ exhibit distinct behaviors when thermal effects are taken into account, consistent with our previous analysis. From Fig. \ref{fig:C_local} (a), we observe that the photon coherence $C_a$ decreases rapidly as the temperature increases for fixed values of $\eta_a$. The maximum coherence $C_a$ can be achieved at low temperatures. Moreover, lower optical frequencies $\eta_a$ lead to enhanced photon coherence $C_a$ in the low-temperature regime, whereas higher values of $\eta_a$ result in improved coherence at elevated temperatures. Fig. \ref{fig:C_local}(b) reveals that higher temperatures can enhance the matter coherence $C_b$ in cases where $\eta_{a} = 0.2, 0.5, 1$, while a gradual decrease in $C_b$ is observed for $\eta_{a} = 5, 10$. This suggests that smaller optical frequencies combined with higher temperatures can enhance the matter coherence $C_b$. In Figs. \ref{fig:C_local}(c) and (d), it is evident that photon coherence $C_a$ decreases as temperature rises, whereas matter coherence $C_b$ increases. Furthermore, with the increase of coupling strengths, the coherences $C_a$ and $C_b$ increase. Figs. \ref{fig:C_local}(e) and (f) demonstrate that the total coherence $C_{\mathrm{tot}}$ displays nonmonotonic behavior, characterized by a dip with rising temperature for $\eta_a = 0.2, 0.5, 1$ when $g/\omega_b=2 $. This dip is also observed in the resonant case for $g/\omega_b = 1, 2$. However, for optical frequencies of $\eta_a = 5, 10$, or for coupling strengths $g/\omega_b = 0.5, 0.1, 0.05, 0.01$, this dip disappears, with the maximum quantum coherence occurring at zero temperature.
Additionally, the quantum coherences $C_b$ and $C_{\mathrm{tot}}$ remain robust with changes in temperature. To further investigate the generation mechanism of quantum coherences, we analyze the contribution of terms $H_{\mathrm{res}}$ and $H_{\mathrm{anti}}$ in the coherence generation, as illustrated in Fig. \ref{fig:bb}. As illustrated in the figure, the total coherence can be enhanced by the term $H_{\mathrm{res}}$, however, it appears to exert no effect on $C_n$. 
Moreover, when considering the term $H_{\mathrm{anti}}$, the total coherence $C_{\mathrm{tot}}$ gradually decreases for $g=0.1, 0.5 \omega_{b}$, however, $C_{\mathrm{tot}}$ increases as the coupling strength becomes stronger with $g/\omega_{b}=1, 2$. As the coupling strengths increase, the total coherence also becomes more pronounced. However, the term $H_{\mathrm{anti}}$ may lead to different coherences $C_a$ and $C_b$. Hence, the total coherence $C_{\mathrm{tot}}$ can be larger than the coherences $C_n$ in the ultrastrong and deep strong coupling regimes. From this perspective, the quantum coherence is different from quantum entanglement as shown in Fig. \ref{fig:EN}.

\begin{figure}
\subfigure[]{
\includegraphics[width=0.35\columnwidth]{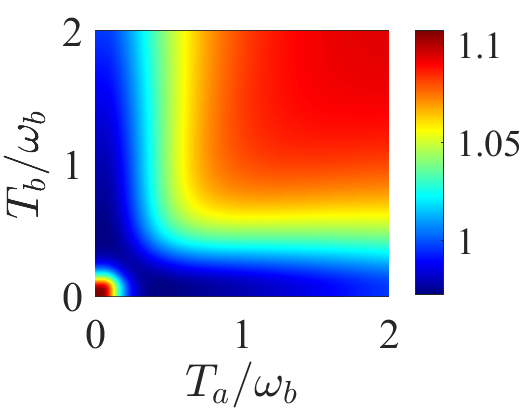}}
\subfigure[]{
\includegraphics[width=0.35\columnwidth]{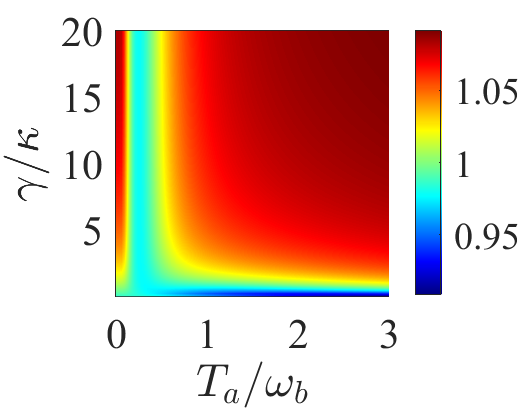}}
\caption{\label{fig:C_3D}
 Quantum coherence $C_{\mathrm{tot}}$ versus (a) the temperatures $T_{a}$ and $T_b$; (b) the temperature $T_L$ and dissipation strength $\gamma$. Note that $\eta_{a}=1$ denotes the resonant case. In (a), the dissipation strength $\gamma=\kappa=10^{-3}$; In (b), $\kappa=10^{-3}$. The values of the parameter can take $g=\omega_b$, and $\eta_a=1$. }
\end{figure}

Here we present a comprehensive analysis of the effect of temperature bias $T_{a}\neq T_b$ and system-reservoir coupling strength on total quantum coherence. Our focus is on the resonant case $\eta_a=1$ and the deep strong coupling regime $g/\omega_b=1$. As illustrated in Fig. \ref{fig:C_3D} (a), our results indicate that maximal coherence is achieved within a low-temperature range. Moreover, we identify a local minimum of coherence at $T_{a}/\omega_b\sim 0.5$. Fig. \ref{fig:C_3D} (b) presents the total coherence as a function of temperature $T_a$ and dissipation strength $\gamma$. Our results indicate that maximal coherence is achieved when $\gamma>\kappa$, either at sufficiently low temperatures or at elevated temperatures.

Furthermore, we discuss the experimental realizations of the current proposals. Recent experimental studies have demonstrated that the coupling of two bosonic modes can be achieved in the ultrastrong and deep strong coupling regimes \cite{WOS:000599002700001, PhysRevApplied.20.024039, PhysRevB.97.024109, WOS:000416229300001, WOS:000558140400001, PhysRevApplied.20.024039, WOS:000554831500031, WOS:000554831500031, WOS:000554831500031, WOS:000599002700001, PhysRevApplied.16.034029, PhysRevLett.120.183601}. Specifically, the ultrastrong coupling of two bosonic modes can be modelled using a variety of interaction forms, such as plasmon-phonon interactions using epsilon-near-zero nanocavities filled with a specific polar medium \cite{WOS:000599002700001}, interacting magnons and photons \cite{PhysRevApplied.20.024039}, coupled mechanical oscillators and electrical circuits \cite{PhysRevB.97.024109},  and three-dimensional crystals composed of plasmonic nanoparticles \cite{WOS:000554831500031}. Moreover, the detection of quantum coherence is accomplished by encoding the coherence in the covariance matrix of the coupled light-matter system. Therefore, one can employ homodyne or heterodyne detection schemes to observe this coherence, similar to the methods discussed in Refs. \cite{PhysRevLett.98.030405, kotler2021direct, WOS:000669369200020}.

%%%%%%%%%%%%%%%%%%%
%%%%%%%%%%%%%%%%%%%

\section{Conclusion}
\label{conclusion}

In conclusion, we have explored the quantum coherences of the photon mode, matter mode, and total system under strong, ultrastrong and deep strong coupling regimes of two bosonic modes. It has been demonstrated that in the ground state, the coherences of the photon and matter modes are equal. Additionally, the coherences for subsystems and total system can be achieved in lower optical frequency ranges and the ultrastrong and deep strong coupling regimes. The coherences $C_{\mathrm{tot}}$ and $C_{n}$ arise from the term $H_{\mathrm{anti}}$, whereas only the term $H_{\mathrm{res}}$ does not contribute to the generation of the coherences. In thermal environments, the term $H_{\mathrm{res}}$ contributes to enhancing total coherence, however, it does not generate $C_n$. Moreover, the total coherence can be enhanced as the coupling strength increases owing to the combined effects of the terms $H_{\mathrm{anti}}$ and $H_{\mathrm{res}}$. Regarding the coherence measure $C_a$, low temperatures yield maximal coherence, while higher optical frequencies facilitate the maintenance of coherence at elevated temperatures. Conversely, lower optical frequencies and higher temperatures can enhance matter coherence $C_b$.
In terms of total coherence, it is observed that lower optical frequencies combined with higher temperatures promote maximal coherence in the deep-strong coupling regime.
Therefore, the ultrastrong and deep strong coupling regimes provide an efficient means to exhibit significant quantum coherences, regardless of whether we are considering the photon mode, matter mode, and total system.

\section*{Acknowledgement}
This work is supported by the China Postdoctoral Science Foundation under Grant No. 2024M760829, the National Natural Science Foundation of China under Grants No. 12175029 and No. 12205193, and the Jiangxi Natural Science Foundation, under Grant No. 20242BAB20035.

\section*{DATA AVAILABILITY}

All datasets that support the findings
of this study are included within the article.

\section*{Appendix}
\setcounter{section}{0}
\setcounter{equation}{0}
\setcounter{figure}{0}
\setcounter{table}{0}
\renewcommand{\theequation}{A\arabic{equation}}
\renewcommand{\thefigure}{A\arabic{figure}}
%\appendix

\section{The coupling photon and matter without $D$ } \label{Appendix A}

\begin{figure}
\subfigure[]{
\includegraphics[width=0.35\columnwidth]{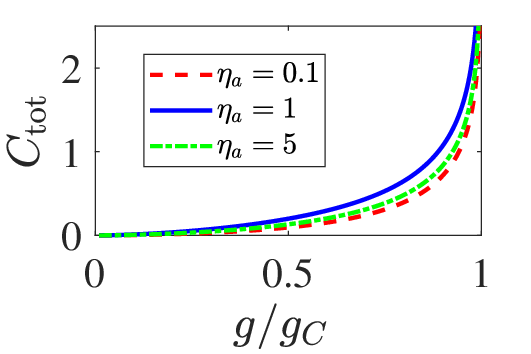}}
\subfigure[]{
\includegraphics[width=0.35\columnwidth]{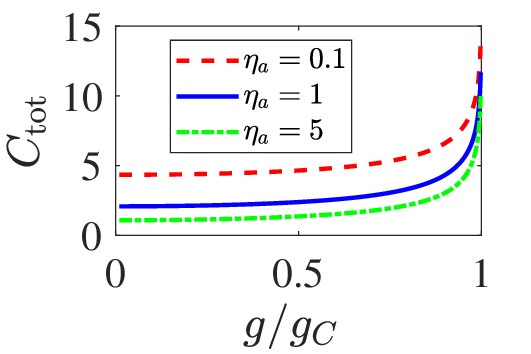}}
\caption{\label{fig:C_No_gs}
Quantum coherence versus the coupling strength $g$ in the ground state (a) and heat environments (b). Note that $\eta_{a}=1$ denotes the resonant case. In (b), the parameters can take $T_{a}=T_{b}=T$ with $T=\omega_b$, and $\gamma=\kappa=10^{-3}$. }
\end{figure}

In the absence of the $D$ term, the coupled light-matter Hamiltonian can be regarded as analogous to a Dicke-type Hamiltonian that exhibits a quantum phase transition at a critical coupling. Specifically, it can be written as
\begin{equation} \label{H_Ss}
H_{S} =\omega_{a} a^{\dagger} a+\omega_{b} b^{\dagger} b+\mathrm{i} g (a+a^{\dagger})(b-b^{\dagger}).
\end{equation}
The polariton frequencies $\omega_{\pm}$ can be obtained by diagonalizing the system Hamiltonian with Eq. (\ref{H_Ss}) as
$\omega_{\pm}=\sqrt{\frac{\omega_a^2+\omega_b^2\pm \sqrt{(\omega_a^2-\omega_b^2)^2+16 g^2 \omega_a \omega_b}}{2}}$. The coefficients of the polariton operators are given by $w_{\pm}=\mathrm{i}\frac{\omega_{\pm}^2+\omega_a \omega_b +(\omega_a+\omega_b) \omega_{\pm}}{2 N_{\pm} g \omega_a}$, $x_{\pm}=\frac{\omega_{\pm}+\omega_b}{N_{\pm}(\omega_{\pm}-\omega_b)}$, $y_{\pm}=\mathrm{i}\frac{\omega_{\pm}^2-\omega_a \omega_b -(\omega_a-\omega_b) \omega_{\pm}}{2 N_{\pm} g \omega_a}$, and $z_{\pm}=\frac{1}{N_{\pm}}$ with $N_{\pm}$ representing the corresponding normalized factors. The critical coupling is determined as $g_C=\sqrt{\omega_a \omega_b}/2$. At the critical coupling $g_C$, the average occupations $\langle a^{\dagger} a \rangle$ and $\langle b^{\dagger} b \rangle$ are diverge, i.e., $\langle a^{\dagger} a \rangle \rightarrow \infty$ and $\langle b^{\dagger} b \rangle \rightarrow \infty$, indicating the emergence of a macroscopic occupation \cite{HEPP1973360, PhysRevB.100.121109}. Consequently, the diamagnetic term plays a crucial role in maintaining the system's stability. In this case, the total quantum coherence increases with the coupling strength prior to reaching the critical coupling $g_C$ as shown in Figs. \ref{fig:C_No_gs} (a) and \ref{fig:C_No_gs} (b). Furthermore, our analysis demonstrates that total quantum coherence diverges in the critical coupling regime. In addition, these critical systems may present promising avenues for advancing quantum metrology \cite{PhysRevE.93.052118, PhysRevResearch.4.043061} and offer a framework for discriminating among distinct master equations \cite{PhysRevResearch.4.013171}.

%%%%%%%%%%%%%%%%%%%%%%%%%%%%%%%%%%%%%%%%%%%%%%%%%%%%%%%%%%%%%%%%%%%%%%%%%%%%%%%%%%%%
%%%%%%%%%%%%%%%%%%%%%%%%%%%%%%%%%%%%%%%%%%%%%%%%%%%%%%%%%%%%%%%%%%%%%%%%%%%%%%%%%%%%

%%%%%%%%%%%%%%%%%%%%%%%%%%%%%%%%%%%%%%%%%%%%%%%%%%%%%%%%%%%%%%%%%%%%%%%%%%%%%
\begin{figure}
\includegraphics[width=0.7\columnwidth]{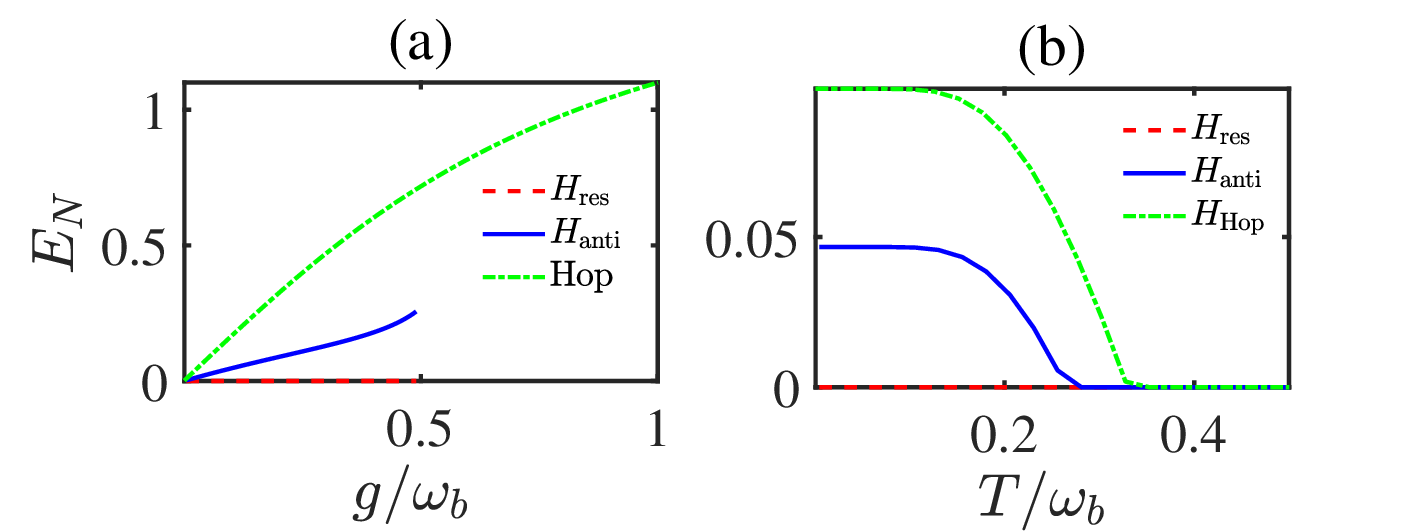}
\caption{\label{fig:EN}
Quantum entanglement $E_N$ versus the coupling strength $g$ in the ground state (a) and heat environment (b). In both figures, the red dashed, blue solid, and green dash-dotted correspond to the full Hamilton without the $H_{\mathrm{anti}}$, without the $H_{\mathrm{res}}$, and full Hamiltonian, respectively. The parameters can take $\eta_a=1$ and $\gamma=\kappa=10^{-3}$ as well as the coupling strength $g=0.1 \omega_b$ for (b). }
\end{figure}
%%%%%%%%%%%%%%%%%%%%%%%%%%%%%%%%%%%%%%%%%%%%%%%%%%%%%%%%%%%%%%%%%%%%%%%

\section{The quantum entanglement } \label{Appendix B}
%%%%%%
In continuous variable systems, quantum entanglement can be quantified using the logarithmic negativity \cite{PhysRevLett.87.167904, PhysRevA.65.032314}. In this framework, the symplectic eigenvalue of the partial transpose of the covariance matrix $\sigma$, denoted by $\tilde{d}_{-}$, is given by $\tilde{d}_{-}=\sqrt{\frac{\Delta - \sqrt{\Delta^2-4 \mathrm{det}\sigma}}{2}}$], where $\Delta=\mathrm{det} \sigma_a+\mathrm{det} \sigma_b-2 \mathrm{det} \sigma_{ab}$.
As illustrated in Fig. \ref{fig:EN}, it is demonstrated that quantum entanglement is primarily generated by the contribution of the term $H_{\mathrm{anti}}$.
The Hamiltonian term $H_{\mathrm{res}}$ is incapable of generating entanglement independently; rather, it enhances entanglement when coupled with the term $H_{\mathrm{anti}}$. Moreover, as the temperature increases, the quantum entanglement diminishes.

\bibliography{coh}

\end{document}